\documentclass[12pt]{iopart}

\usepackage{graphicx}

\begin{document}

\title{Studies of reversible capsid shell growth}

\author{D. C. Rapaport}

\address{Department of Physics, Bar-Ilan University, Ramat-Gan 52900, Israel}

\ead{rapaport@mail.biu.ac.il}

\date{September 06, 2009}

\begin{abstract}

Results from molecular dynamics simulations of simple, structured particles
capable of self-assembling into polyhedral shells are described. The analysis
focuses on the growth histories of individual shells in the presence of an
explicit solvent and the nature of the events along their growth pathways; the
results provide further evidence of the importance of reversibility in the
assembly process. The underlying goal of this approach is the modeling of virus
capsid growth, a phenomenon at the submicroscopic scale that, despite its
importance, is little understood.

\end{abstract}


\section{Introduction}

The growth of viral capsids -- the polyhedral shells of capsomer particles
enclosing the genetic package of spherical viruses \cite{cri56,cas62} -- is one
of the more familiar examples of supramolecular self-assembly. The fact that
assembly also occurs {\em in vitro}, in the absence of genetic material
\cite{pre93,zlo99,cas04}, simplifies the overall assembly problem and makes it
an ideal candidate for simulation. A further reduction in complexity arises from
the fact that icosahedral symmetry is ubiquitous among spherical viruses, where
capsid shells are formed from an appropriate number of copies of just one or a
small number of distinct capsomers; this implies that as an initial
approximation, the molecular details of the capsomer proteins can be expressed in
a highly reduced, nonspecific form. Molecular dynamics (MD) simulation employing
simplified models of this kind ought to be capable of examining the existence
and nature of universal organizational principles governing capsid
self-assembly.

Simulation has an especially important role in the study of assembly pathways
given that nonequilibrium systems are involved and, as a consequence, very
little direct experimental evidence is available. Refs.~\cite{rap99,rap04}
describe MD modeling of capsid self-assembly based on simplified structural
models in which the particle representation retains sufficient detail to ensure
meaningful behaviour. The principal characteristics of the model are (a) an
effective molecular shape formed out of rigidly packed soft spheres that enables
particles to fit together in a closed shell, and (b) multiple interaction sites
located to ensure that the minimal-energy structures, both intermediate and
final, have the desired forms. The pathways themselves were not considered in
the initial work, since the emphasis was on demonstrating the feasibility of
assembly, and computational limitations required omission of an explicit
solvent.

In a more recent study \cite{rap08}, self-assembly in the presence of an
explicit atomistic solvent was described. Again there were computational
limitations, in this case the limitation to icosahedral shells constructed from
triangular particles, rather than the previously considered shells of size 60,
and larger, formed from more elaborate trapezoidal particles, but there is no
reason to question the generality of the observed behaviour. The presence of a
solvent aids the breakup of partially assembled shells without subassemblies
needing to collide directly, weakens the ballistic contribution to particle
movement, and serves as a heat bath to absorb energy released during exothermal
bond formation while helping to ensure conditions closer to thermal equilibrium.
The results described in Ref.~\cite{rap08} lead to the conclusion that
self-assembly consists of a cascade of reversible stages, in which low-energy,
maximally bonded intermediate states are strongly preferred, a process that
eventually yields a high proportion of completely assembled shells. Despite the
apparent paradox, the efficiency of the overall assembly process depends on
reversibility, one of whose contributions is to help avoid the consequences of
trapped states.

There have been other studies that address the dynamics of capsid assembly. An
alternative particle-based, solvent-free MD simulation involved quasi-rigid
bodies formed from hard spheres \cite{ngu07}. More highly simplified capsomer
representations have been based on spherical particles, instead of extended
capsid shapes, with either directional interactions \cite{hag06} whose range
exceeds the particle size, or bonding energies determined by local neighborhood
rules \cite{sch98}; in these simulations the solvent is represented implicitly
by stochastic forces. A further alternative involves Monte Carlo simulation of
patchy spheres \cite{wil07}, but here the dynamics of assembly are of course
absent. At the other extreme on the complexity scale are the folded proteins of
real capsomers, although MD simulations employing all-atom models \cite{fre06}
are limited to short trajectories for testing the stability of prebuilt shells.
A variety of theoretical methods have also been harnessed to study capsid
structure \cite{lid03,twa04,zan04,hic06,hem06}, while concentration kinetics
have been used for interpreting experiments \cite{zlo99,van07}; discrete
particle dynamics are not involved in such approaches.

The present paper extends the analysis of the simulations initially described in
Ref.~\cite{rap08}. An alternative approach to probing the evolution of partially
assembled structures will be introduced that is able to provide additional
details about events occurring along the assembly pathways. The method is based
on establishing the identity of each partial assembly at any given stage on the
pathway by referring to the complete shell that eventually forms containing a
majority of its current member particles. This permits monitoring the evolution
of individual clusters of particles as each develops into a closed shell,
allowing for addition and loss of members along the pathway.

\section{Methodology}

\begin{figure}
\begin{center}
\includegraphics[scale=0.50]{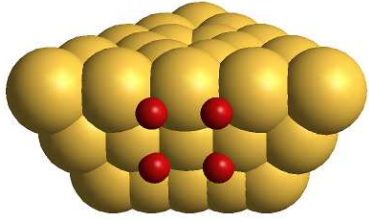}
\end{center}
\caption{\label{fig:1} (Color online) Model particle showing arrangement of
spheres that determines overall shape; small spheres denote locations of
interaction sites.}
\end{figure}

The simulations \cite{rap08} involve particles whose effective shape is the
truncated triangular pyramid shown in Figure~\ref{fig:1} designed to
self-assemble into icosahedral shells. The larger, slightly overlapping spheres
that provide the overall shape occupy multiple planes, while the interaction
sites, represented by small spheres for visual convenience, determine the
locations and orientations of the the three lateral faces; these are inclined at
$20.905^\circ$ to the normal. Each lateral face contains four interaction sites
that can bond to matching sites on adjacent particles; the reason for multiple
sites is that they help maintain correct alignment, a feature that is
particularly important for partial assemblies in which particle attachment is
incomplete. The particle  structure and interactions are based on the model
introduced in Ref.~\cite{rap04}. This highly simplified representation can be
contrasted with real capsomers \cite{bak99} that consist of intricately folded
proteins whose exposed surfaces form relatively complex landscapes.

The same spheres responsible for the shape of the particle also represent the
solvent atoms. All spheres experience a (short-range) soft-sphere repulsion
based on the truncated Lennard-Jones potential; the parameters of the potential
determine the dimensionless MD length and time units \cite{rap04bk} used in the
simulations. The attractive force between bond-forming interaction sites is
derived from the potential
\begin{equation}
u(r) = \left\{
\begin{array}{ll}
e (1 / r_a^2 + r^2 / r_h^4 - 2 / r_h^2) & \quad r < r_h \\[4pt]
e (1 / r_a^2 - 1 / r^2) & \quad r_h \le r < r_a
\end{array}
\right.
\end{equation}
This interaction is harmonic at distances below $r_h=0.3$ and inverse-square
above $r_h$, with range $r_a=3$; its overall strength is governed by the
parameter $e$ that distinguishes the different runs described below. Particle
size exceeds the interaction range, although less so than in real capsomers; the
effect is to reduce the attraction between wrongly positioned or oriented
particles. Other more general aspects of MD methodology, including the
interaction computations and integration of the equations of motion, are
described in Ref.~\cite{rap04bk}.

The system consists of 1875 triangular particles, sufficient for producing 93
full shells; there are a total of 125\,000 molecules, the majority (98.5\%) of
which are solvent atoms. The system is confined to a cubic region with periodic
boundaries; region size is chosen to yield a mean number density of 0.2. Even
though the particle concentration is much higher than in experiment, the solvent
presence is adequate to ensure that diffusion minimizes the effects of the
ballistic particle motion that would otherwise occur. The run length is $60
\times 10^6$ time steps, adequate for ensuring that essentially all growth has
ceased; 200 steps correspond to one unit of (MD) time. Particle mass is set at
21 times that of the solvent atom (with unit mass); having a much smaller mass
ratio than in reality shortens the assembly timescale, making it accessible to
MD, but without altering the behaviour in any qualitatively significant manner.

Bond formation is exothermal and leads to a gradual warming of the system; this
is suppressed by means of a thermostat that maintains a temperature
corresponding to unit mean (translational and rotational) kinetic energy per
particle. In the initial state, particles and solvent atoms are positioned on a
lattice with random velocities; to avoid possible overlap at the start,
particles begin collapsed (with all their component spheres fully overlapped)
and expand to their final shape over the initial 5000 steps. None of the
additional mechanisms that were introduced to regulate or assist assembly (e.g.,
damping, or the breakup of partial assemblies) described in Ref.~\cite{rap04}
are necessary for the present simulations.

Establishing membership of partial assemblies and algorithmically verifying that
shells are correctly assembled requires the capability for identifying bound
clusters \cite{rap04bk} and checking the connectivity of their bond networks.
Cluster membership is a key part of the analysis and, in the present study,
interaction sites are regarded as bonded when less than 0.6 ($= 2 r_h$) apart;
this threshold is empirically chosen to avoid transient apparent bond breakage
caused by thermal vibration. The particles themselves are considered bonded if
all four site pairs on the adjoining faces are bonded; this state implies almost
complete particle alignment, with only the smallest of fluctuations in relative
position and orientation.

\section{Results}

The analysis begins with a comparison of the different growth scenarios observed
as the interaction strength parameter $e$ is varied; if the range of variation
is not too large this is equivalent to examining the temperature dependence of
the behaviour. A more detailed discussion of shell growth for the maximal yield
case then follows.

Table~\ref{tab:1} summarizes the outcome of a series of simulation runs for
various values of $e$; the results are expressed as the mass fraction contained
in clusters of different sizes and the residual particles present as monomers.
Essentially no change in cluster population occurs towards the end of the runs.
At low $e$ very little growth occurs due to minimal initiation, but as $e$ is
increased the balance shifts towards higher yields of complete shells. The
maximum yield of 83 shells (out of a possible 93) is achieved at $e=0.13$. The
yield then falls, since the ability to reach completion is inhibited by
excessive early growth, resulting in too many monomers being incorporated into
clusters prematurely. Repetition of one of the runs with a different initial
state confirmed that, allowing for reasonable fluctuations, the results are
reproducible. No oversized (mutant) clusters appeared in these runs, although
these would be expected for sufficiently large $e$.

\begin{table}
\caption{\label{tab:1} Final cluster distributions for different interaction
strengths $e$; mass fractions of monomers, clusters grouped by size range, and
complete shells are listed, with the maximum mass fraction for each run shown
in bold.}
\begin{center}
\begin{tabular}{lcccccc}
\\
\hline
\\
 $e$    &       \multicolumn{6}{c}{Cluster mass fraction}      \\
        &Size: 1 & 2-5     & 6-10   & 11-15  & 16-19  & 20     \\
\\
\hline
\\
 0.11   & \textbf{0.7931} & 0.0976  & 0.0181 & 0.0080 & 0.0085 & 0.0747 \\
 0.115  & \textbf{0.5153} & 0.0704  & 0.0053 & 0.0256 & 0.0101 & 0.3733 \\
 0.12   & 0.3040 & 0.0314  & 0.0032 & 0.0000 & 0.0000 & \textbf{0.6614} \\
 0.125  & 0.1915 & 0.0283  & 0.0032 & 0.0203 & 0.0101 & \textbf{0.7466} \\
 0.13   & 0.0709 & 0.0182  & 0.0032 & 0.0224 & 0.0000 & \textbf{0.8853} \\
 0.14   & 0.0011 & 0.0000  & 0.0310 & 0.1104 & 0.2282 & \textbf{0.6293} \\
 0.15   & 0.0000 & 0.0000  & 0.0192 & 0.3158 & \textbf{0.4623} & 0.2027 \\
\\
\hline                                                                   
\end{tabular}
\end{center}
\end{table}

\begin{figure}
\begin{center}
\includegraphics[scale=0.95]{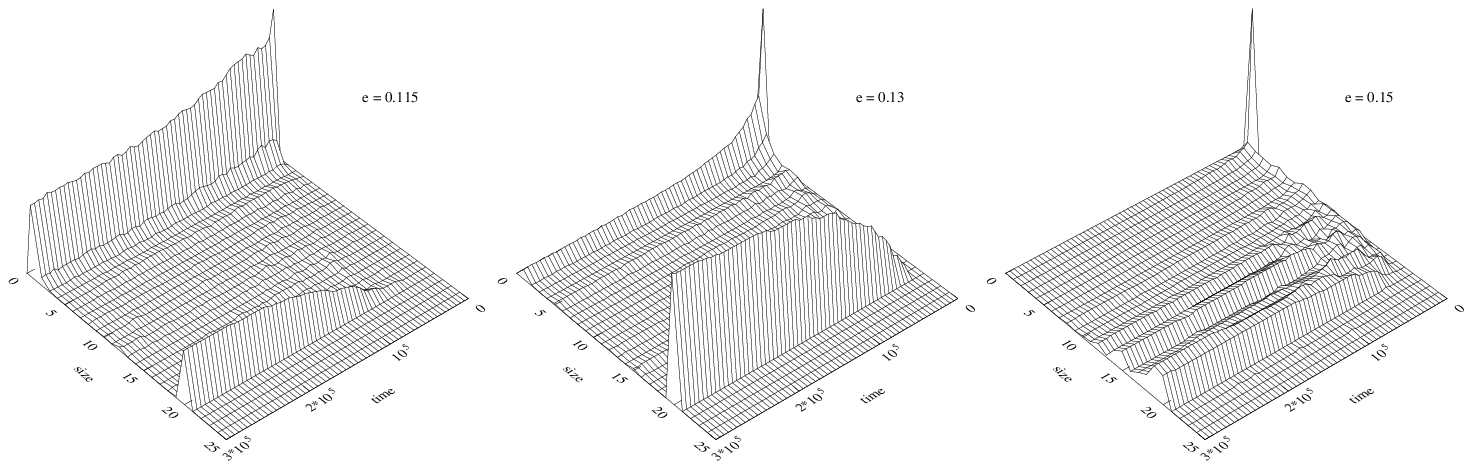}
\end{center}
\caption{\label{fig:2} Cluster size distributions (including monomers) as
functions of time (MD units) for different attraction strengths $e$; the
distributions are expressed as mass fractions.}
\end{figure}

\begin{figure}
\begin{center}
\includegraphics[scale=0.25]{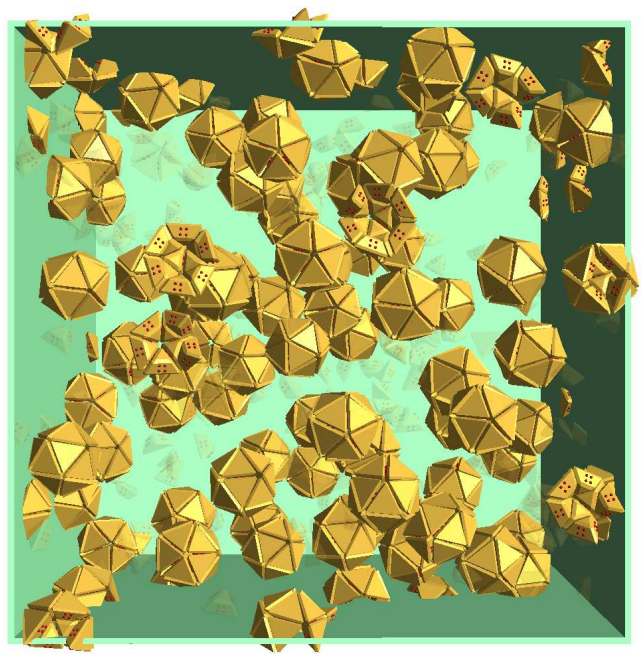}
\end{center}
\caption{\label{fig:3} (Color online) Image of the $e=0.13$ system at the end of
the run; the relatively few particles not in complete shells are shown
semi-transparently and the solvent is omitted (visual artifacts due to periodic
boundaries are mentioned in the text).}
\end{figure}

The time development of the cluster size distributions, also expressed in terms
of mass fractions, is shown in Figure~\ref{fig:2}. This is a subset of the
results shown in Ref.~\cite{rap08}, but the plots are shown from a different
perspective to allow the early and intermediate growth features, especially the
limited population of small clusters, to be seen more clearly.
Figure~\ref{fig:3} shows an image of the $e=0.13$ system once all 83 complete
icosahedral shells have formed, with other partial structures shown ghosted.
Note that periodic boundaries are applied at the level of individual particles,
so that shells that cross the container boundaries appear fragmented; while the
solvent particles are not shown here (for clarity, unlike \cite{rap08}) they
actually fill the volume. As described in Ref.~\cite{rap08}, closed shells are
especially stable, so that even if $e$ is subsequently reduced to a value too
low for assembly initiation, the shells do not self-destruct, implying
hysteresis.

The earlier analysis of shell growth \cite{rap08} dealt with the accumulated
lifetimes of particle clusters of different sizes and the probabilities of
events corresponding to size increases and decreases, with the latter tending to
dominate. In addition, the energetics of intermediate structures were
considered, and a strong preference for maximally bonded (low energy)
configurations was observed. The focus was on the mean cluster properties as a
function of size, while the assembly history of individual shells was not
considered; the following discussion addresses this topic.

The examination of the growth history of individual shells begins by considering
the shell membership of the final state of the run. On the basis of this
information, as indicated above, it is possible to associate partial assemblies
at earlier stages of the run with particular final shells, based on the majority
membership of their particles. In general, this accommodates particles both
entering and departing the growing cluster. There can be some identity
ambiguity, however, for small clusters, where a given final shell can own the
majority of particles from more than one such cluster, or the identity of the
cluster with the most particles destined for a given shell can change. Such
effects will have only minimal influence on the ability to monitor individual
cluster histories once a relatively stable core (e.g., a pentagon) has formed,
and they become even less of a concern as growth progresses further.

The analysis of the properties of a given cluster considers all its member
particles, including those that subsequently detach and do not belong to the
majority. Cluster membership is determined from configurations recorded every
2000 time steps (10 MD time units); the limited time resolution can merge or
conceal multiple closely spaced events, but in view of the relatively slow
particle motion, the majority of individual growth steps can be distinguished.
The analysis is based on the high-yield $e=0.13$ system; in addition to the 83
complete shells, there are 11 incomplete and small clusters that are not
considered.

\begin{figure}
\begin{center}
\includegraphics[scale=0.90]{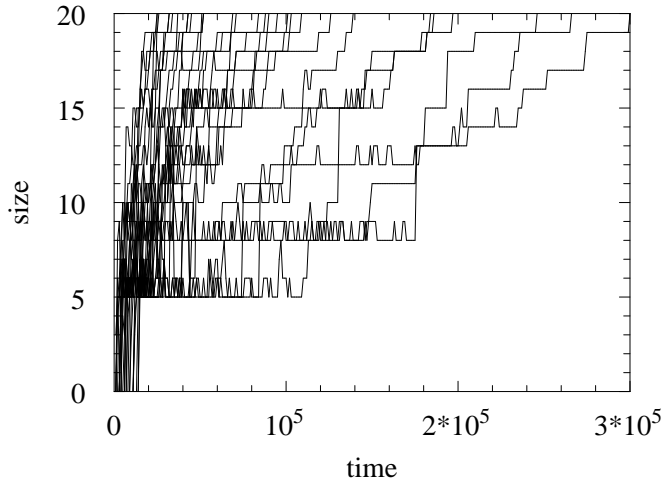}
\end{center}
\caption{\label{fig:4} Individual size histories for 30 out of the 83 shells.}
\end{figure}

The growth histories of a subset (for clarity) of 30 out of the total of 83
shells are shown in Figure~\ref{fig:4}. Initial growth to pentamer size, of
which 95\% are regular pentagons \cite{rap08}, occurs rapidly, but the
distribution of subsequent growth rates is broad. While some clusters grow
rapidly -- some even monotonically -- to completion, the paths of others appear
to become blocked at certain sizes, repeatedly adding and then promptly losing
an additional particle until, eventually, a more lasting growth step is
achieved. Size increases greater than unity are apparent. Analysis in greater
detail, preferably aided by direct visualization, would be needed to determine
if there is any clear distinction, e.g., in terms of detailed morphology,
between clusters experiencing fast and slow growth rates, or whether the growth
histories are dominated by fluctuations.

\begin{figure}
\begin{center}
\includegraphics[scale=0.90]{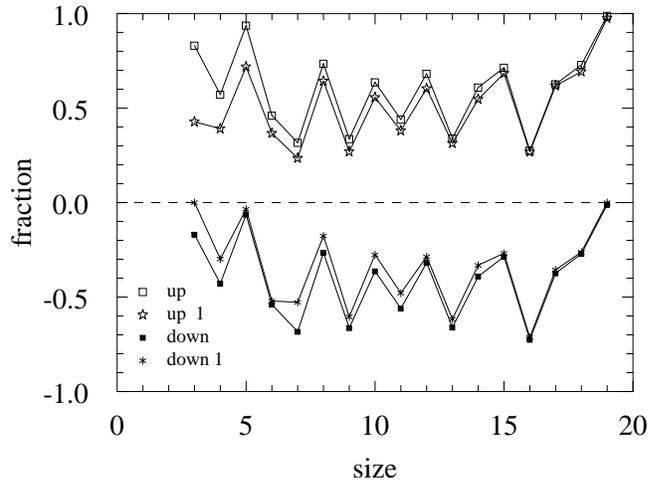}
\end{center}
\caption{\label{fig:5} Fraction of size-change events occurring for clusters of
each size; size-decrease events are shown with negative values for emphasis.}
\end{figure}

Figure~\ref{fig:5} shows the fractions of events corresponding to unit size
changes in each direction, together with the fractions of all size-changing
events irrespective of magnitude, each as a function of cluster size; negative
values are used to distinguish the size-decrease events, so it is the distances
from the zero line that must be compared. For most cluster sizes, unit size
changes account for the majority of events. The important role of reversibility
is clear from these measurements, as already noted in Ref.~\cite{rap08}, namely
that a substantial fraction of events at all sizes (except 5 and 19) involve
size decreases, and that there are cluster sizes (e.g., 7, 9, 13, 16) for which
the size is more likely to decrease than increase.

Additional details emerge from considering the detailed breakdown of size
changes (not shown). From the particular run under examination it is apparent,
for example, that the probability of a given size change does not vary
monotonically with the magnitude of the change, and changes of size $\pm 5$ have
an increased likelihood: for clusters of size 10, 27\% of the events are
reductions of -1, 5\% -2, 0.4\% each -3 and -4, and 4\% -5, whereas for size 12,
60\% of the events are increases of +1, 4\% +2, 1\% +3, 0\% +4, and 2\% +5.

\begin{figure}
\begin{center}
\includegraphics[scale=0.90]{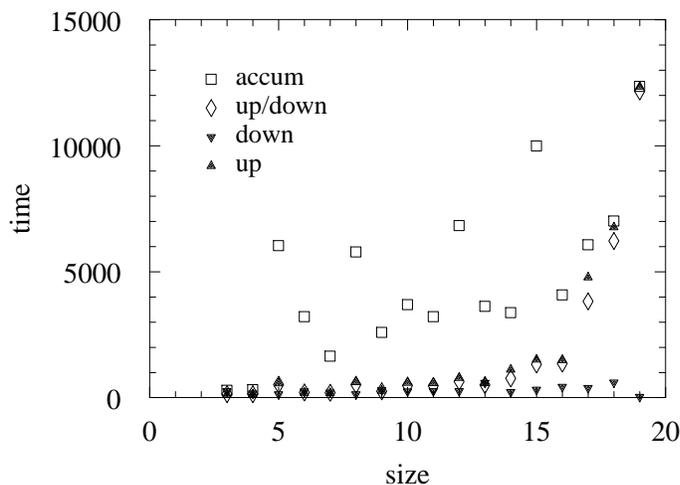}
\end{center}
\caption{\label{fig:6} Measurements of mean cluster lifetimes (see text for
explanation of quantities).}
\end{figure}

Measurements of cluster lifetime appear in Figure~\ref{fig:6}. Four different
quantities are shown. The accumulated time is the mean total time that clusters
exist at a given size (the lifetime distributions themselves are broad). The
values can be correlated with the preferred direction of change, and those sizes
where increases are more likely to occur than decreases (notably 5, 8, 10, 12,
15, and 19) also have large accumulated times. As a consequence of the strongly
reversible nature of cluster growth, the time spent at a particular size is
likely to be made up of several distinct intervals. Thus the second quantity
shown, referred to as up/down, is the mean uninterrupted time spent at a given
size, a value that in most cases is substantially less than the accumulated
time; the ratio of these times provides an estimate of the number of visits to
the corresponding size, and the value ranges from a low of almost unity at size
19, up to about 15 at size 5. The remaining quantities are breakdowns of the
uninterrupted time according to the direction of the next size change; in those
cases where a clear difference exists, it is apparent that the time to wait for
a size decrease can be considerably less than for a size increase. These results
provide additional evidence of the way reversibility dominates the overall
growth process.

\begin{figure}
\begin{center}
\includegraphics[scale=0.90]{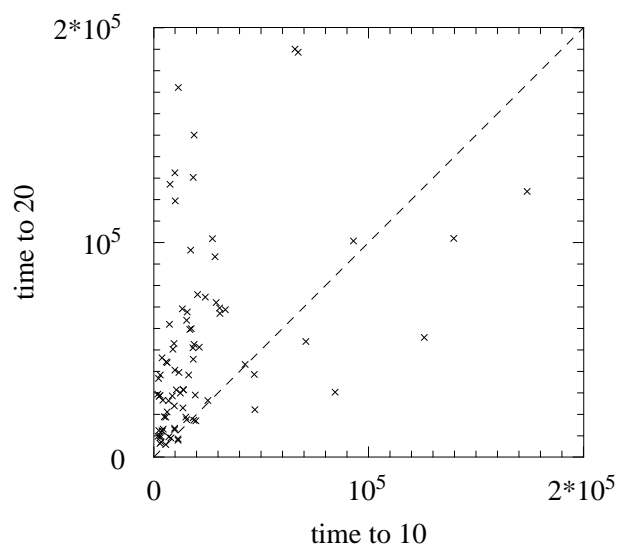}
\end{center}
\caption{\label{fig:7} Scatter plot showing the time to reach size 10, and then
size 20, for each shell.}
\end{figure}

A further characteristic of the growth history is the influence of the rate of
early growth on the speed of subsequent development. The scatter plot shown in
Figure~\ref{fig:7} provides a simple way to examine this effect. Each data point
corresponds to a shell, where the coordinates denote the elapsed time to reach
size 10 and the time from size 10 to completion. For those clusters lying above
the diagonal the latter time is longer, in some cases by a substantial amount,
but for a small proportion of shells it is the first half of the growth process
that is the more time consuming.

\begin{figure}
\begin{center}
\includegraphics[scale=0.70]{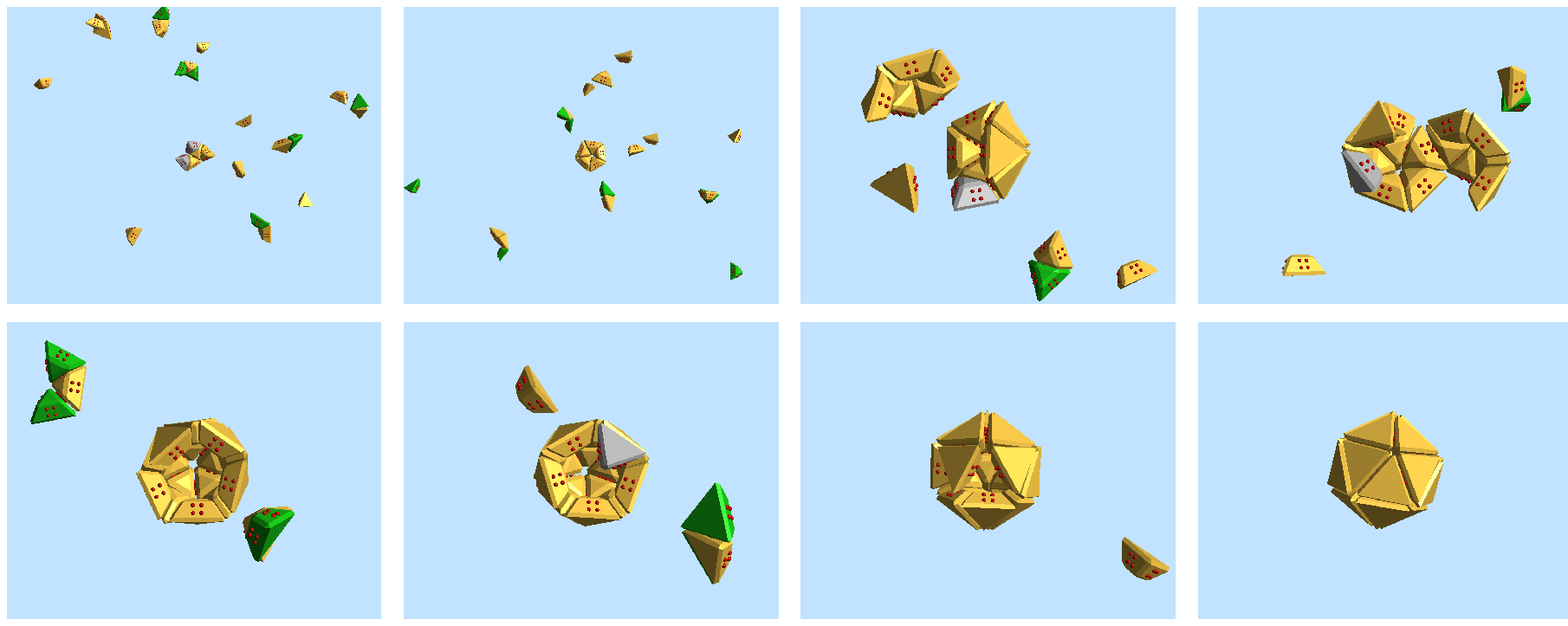}
\end{center}
\caption{\label{fig:8} (Color online) Images showing the growth of one of the
shells; only the particles directly involved are included although some lie
outside the field of view (solvent is also omitted); the color coding is
explained in the text.}
\end{figure}

Imagery is especially helpful for exploring those aspects of the growth process
that are less readily quantifiable, and offers the possibility of suggesting
additional approaches to analyzing the pathway details. Figure~\ref{fig:8} shows
a series of images covering several stages in the growth of just one of the
shells. Only the particles directly involved are included (although some may be
too far away to appear in the frames shown). Color coding identifies the
eventual disposition of the particles; yellow for particles destined for (or
already in) the final shell, gray for particles only temporarily attached to the
growing shell, and green for particles that are temporarily attached to yellow
particles not yet in the final shell. The particular growth sequence shown here
turns out to be an atypical one, based on an analysis of event types, since it
includes the joining of two clusters both of which are at least of pentamer
size. A pentagon is seen in frame \#2, two larger complexes in frame \#3 and a
cluster merging event in \#4, the shell nearing completion in \#7, and the final
complete shell in \#8.

\section{Conclusion}

The present paper continues the study of the dynamics of simplified viral
capsomer models in an explicit atomistic solvent, the eventual goal of which is
the modeling of capsid self-assembly. The current focus is on measurements
related to the growth histories of individual polyhedral shells. This form of
analysis provides an alternative perspective, as well as support for the earlier
results that revealed the importance of reversibility. As the results clearly
show, microscopic self-assembly, where the dynamics reflect the intrinsic
thermal fluctuations prevalent at such scales, is an entirely different class of
phenomenon from corresponding processes at macroscopic scales where
reversibility is not a consideration.

\section*{References}

\bibliography{revcapgrow}

\bibliographystyle{unsrt}

\end{document}